%% file: geosch2pn3.tex
\documentclass[a4paper,english,aps,manuscript]{revtex4}
\usepackage[T1]{fontenc}
\usepackage[latin9]{inputenc}
\usepackage{amsmath}

\makeatletter

\usepackage{amsfonts}

\usepackage{bbm}

\usepackage{amscd}

\usepackage{array}

\usepackage{tensind}

\usepackage{mathrsfs}

\tensordelimiter{?}

\usepackage{longtable}

\usepackage{color}

\def\laq{\raise 0.4ex\hbox{$<$}\kern -0.8em\lower 0.62ex\hbox{$\sim$}}
\def\gaq{\raise 0.4ex\hbox{$>$}\kern -0.7em\lower 0.62ex\hbox{$\sim$}}

\newcommand{\ba}{\begin{array}}
\newcommand{\ea}{\end{array}}

\newcommand{\be}{\begin{equation}}
\newcommand{\ee}{\end{equation}}
\newcommand{\bea}{\begin{eqnarray}}
\newcommand{\eea}{\end{eqnarray}}

\newcommand{\mytextrm}[1]{{}}

\def\ee{{\mathrm{e}}}

\newlength{\sizeonefig}
\newlength{\sizetwofig}
\newlength{\sizeonefigb}
\newlength{\sizetwofigb}
\makeatother

\usepackage{babel}

\newcommand{\kreu}{\times}
\newcommand{\clight}{c}

\newcommand{\op}[1]{\operatorname{#1}}
\newcommand{\fett}[1]{\bold{#1}}
\newcommand{\parenmacro}[1]{\left(#1\right)}

\newcommand{\pqm}{1}
\newcommand{\pqe}{2}

\newcommand{\mme}{m_{\pqm}+m_{\pqe}}

\newcommand{\mm}{m_{\pqm}}
\newcommand{\me}{m_{\pqe}}

\newcommand{\myvec}[1]{\fett{#1}}

\newcommand{\ps}{\myvec{p}}

\newcommand{\xs}{\myvec{x}}

\newcommand{\vpm}{\myvec{p}_{\pqm}}

\newcommand{\vpe}{\myvec{p}_{\pqe}}

\newcommand{\vxm}{\myvec{x}_{\pqm}}

\newcommand{\vxe}{\myvec{x}_{\pqe}}

\newcommand{\xz}{\myvec{X}}

\newcommand{\pzabs}{P}
\newcommand{\pz}{\myvec{P}}

\begin{document}

\title{Canonical center and relative coordinates for compact binary systems
through second post-Newtonian order}

\author{Ira Georg\footnote{grgira001@myuct.ac.za}}

\affiliation{Department of Mathematics and Applied Mathematics, University of Cape Town, Rondebosch, 7701, South Africa}

\author{Gerhard Sch\"afer}

\affiliation{Theoretisch-Physikalisches Institut, Friedrich-Schiller-Universit\"at Jena, Max-Wien-Platz 1, 07743 Jena,
Germany}

\begin{abstract}
Based on a recent paper by Rothe and Sch\"afer on compact binary systems [J. Math. Phys. \textbf{51}, 082501 (2010)],
explicit expressions for canonical center and relative coordinates
in terms of standard canonical coordinates are derived for spinless objects up to second post-Newtonian approximation
of Einstein's theory of gravity (the third post-Newtonan order expressions are available in form of supplementary data). 
The inverse relations, i.e. the dependence of the standard canonical coordinates on the 
canonical center and relative coordinates, are also given up to the second post-Newtonian approximation. 
The famous Pythagorean-theorem-type Lorentz-invariant relation between the system's total energy or Hamiltonian squared, the rest energy or mass squared
-- solely depending on relative coordinates --, and the total linear momentum squared are explicitly shown through second post-Newtonian approximation.     
\end{abstract}


\date{\today}

\maketitle

\section{Introduction}

Compact binaries are important objects in the relativistic astronomy and astrophysics. Their orbital motions however are difficult to calculate if the 
Einstein theory of gravity is deeply needed for sufficient accuracy. Analytically, only approximate solutions are known whereby post-Newtonian (PN) approximation calculations 
have shown to be most efficient, e.g. \cite{CW11}. Very helpful is the knowledge of Hamiltonians because they easily allow the transition to the rest frame by putting the total linear momentum, 
which simply is the linear sum of the linear momenta of the single components, equal to zero which is allowed whenever the center-of-mass motion is unimportant. In this way, the rest-frame dynamics of 
post-Newtonian gravitationally bounded binary systems has often been derived, e.g.\ \cite{DS88,SW93,DJS00,DJS00a,MGS05}. However, if for example a binary system is part of 
a hierarchical triple system its total motion becomes crucial and a more general treatment is required, e.g.\ \cite{AR05,PW14,CW14,RS14}. The same holds if the recoil of merging binary systems through gravitational 
wave emission cannot be neglected, see e.g.\ \cite{LZ11,TBW10}. A specific topic in the gravitational recoil problem is the generalization of the effective-one-body method to the binary-center 
motion, see e.g. \cite{DG06}. These topics clearly make a strong case for the investigations performed in the present paper. 

The transition from standard canonical coordinates to center and relative coordinates is nothing but a canonical transformation. The generators of this 
canonical transformation have been derived up to third post-Newtonian (3PN) order, even including spin degrees of freedom, in \cite{RS10}.
However, explicit expressions for the relations between the center and relative coordinates and the standard canonical coordinates of the binary components, 
and vice versa, have not been given. The present paper closes this gap through 2PN order for spinless binaries by tedious but straightforward calculations 
(the generalization of our Sec. III through 3PN order is available in form of supplementary data; 
the more complicated case of spinning binaries is left for future research). Relying on the diploma thesis by one of the authors, \cite{IG13}, we give explicit expressions for canonical center and 
relative coordinates in terms of standard canonical coordinates up to 2PN order. 
We particularly show that, using these expressions, the Pythagorean-theorem-type Lorentz-invariant relation between 
the system's total energy (or Hamiltonian squared), the rest energy (or mass squared, only depending on relative coordinates), and the total linear momentum 
squared holds up to 2PN approximation. Working within a canonical framework we are always allowed to make canonical transformations. This means that neither the Hamiltonian at any PN order 
nor the various PN coordinate expressions we are calculating are uniquely defined in the quasi-Minkowskian coordinates that we are using throughout our paper (recall the Schwarzschild metric in 
isotropic coordinates or in Schwarzschild coordinates but with cartesian components). Thus, no local physical interpretation of our PN expessions can be given.
Readers interested in historical information on the relativistic two-body problem in relation to center and relative coordinates may find details in 
e.g.\ \cite{AC07}.

\section{Fundamentals}
We start with standard canonical position and linear momentum coordinates (variables) for the components of a binary system, denoted by $(\myvec{x}_a,\myvec{p}_a)$, with coordinate-components $(x^i_a = x_{ai}, p_{ai} = p^i_a)$ -- using four-dimensional metric signature +2, we do not need to discriminate between covariant and contravariant components of our cartesian coordinates --, where the labels $a = (1,2)$ and $i = (1,2,3)$ number the involved objects and the coordinate-components, respectively. These variables are called the same in all our various PN approximations. Also, the total linear momentum, ${\bf P} = \sum_a {\bf p}_a$, and the total orbital angular momentum, ${\bf L} = \sum_a {\bf x}_a \times {\bf p}_a$, are given by the same expressions through all PN orders. 
A simple transformation takes us to the following set of coordinates, which we call Newtonian coordinates, because in the realm of Newtonian physics they are nothing but the Newtonian center-of-mass, $\xz^{\op{N}}$, the total linear momentum, $\pz^{\op{N}} (= {\bf P})$, the relative position, $\xs^{\op{N}}$, and the relative momentum, $\ps^{\op{N}}$, coordinates of two objects with mass parameters $m_a$, 
\begin{align}
  \xz^{\op{N}}  & =  \frac{\mm}{\mme}\vxm+\frac{\me}{\mme}\vxe,
  &
  \xs^{\op{N}}  & = \vxm-\vxe,
  \nonumber \\
  \pz^{\op{N}}  & = \vpm+\vpe,
  &
  \ps^{\op{N}}  & = \frac{\me}{\mme}\vpm-\frac{\mm}{\mme}\vpe.
\end{align}

Within the Hamiltonian framework a set of coordinates is called canonical if the equations of motion are in Hamiltonian form, 
\begin{align}
 \dot{x}^i_a &= \frac{\partial H}{\partial p^{\,i}_{a}} = \{x^i_a,H\}, &   \dot{p}^i_a &= -\frac{\partial H}{\partial x^{\,i}_a} = \{p^i_a,H\} ,
\end{align}
where $\{,\}$ denotes the Poisson brackets.

Every set of canonical coordinates can be used to evaluate the Poisson brackets and the canonical coordinates satisfy the canonical commutation relations, which read for the two different sets of canonical coordinates,
\begin{align}
 \left\{x_{ai},\, p_{bj}\right\} =\delta_{ab}\delta_{ij}\, ,\quad \hbox{zero otherwise},\\
 \left\{X^{\rm{N}}_{\,i},\, P^{\rm{N}}_{\,j}\right\} = \delta_{ij}\, ,\quad
 \left\{x^{\rm{N}}_{\,i},\, p^{\rm{N}}_{\,j}\right\} = \delta_{ij}\, ,\quad \hbox{zero otherwise}.
\end{align}

We are looking for a new set of canonical coordinates ($\mathbf{X},\mathbf{P},\mathbf{x},\mathbf{p}$) such that the equations of motion decouple into an external and an internal part. In Newtonian physics we simply insert a Newtonian Hamiltonian ($H = \frac{{\bf p}_1^2}{2m_1} + \frac{{\bf p}_2^2}{2m_2} + V(|{\bf x}_1-{\bf x}_2|)$) and find that the equations decouple, 
\begin{align}
 \dot{X}^{\op{N}}_{\,i} &= \{X^{\op{N}}_{\,i},H\} = \frac{\partial H}{\partial P^{\op{N}i}} = \frac{P^{\op{N}}_{\,i}}{m} \,, &
 \dot{x}^{\op{N}}_{\,i} &= \{x^{\op{N}}_{\,i},H\} = \frac{\partial H}{\partial p^{\op{N}i}} = \frac{p^{\op{N}}_{\,i}}{\mu}  ,\nonumber\\
 \dot{P}^{\op{N}}_{\,i} &= \{P^{\op{N}}_{\,i},H\} =-\frac{\partial H}{\partial X^{\op{N}i}} = 0 \,, &
 \dot{p}^{\op{N}}_{\,i} &= \{p^{\op{N}}_{\,i},H\} =-\frac{\partial H}{\partial x^{\op{N}i}} = -\frac{G\mu m}{r_{12}^3} x^{\op{N}}_{\,i},
 \label{NewtonianDecoupling}
\end{align}
where $m = m_1 + m_2$, $\mu = m_1m_2/m$, and $r_{12} = |{\bf x}_1-{\bf x}_2|$. Apparently, the quantities that determine the external motion (${\xz}^{\rm N},{\pz}^{\rm N}$) are independent from those quantities of the internal motion (${\xs}^{\rm N},{\ps}^{\rm N}$) and vice versa.

In the relativistic theory we have to make an effort to decouple the equations of motion. We note that the energy-momentum relation, with $c$ denoting the speed of light,
\begin{align}\label{energymomentumequation}
 E^2 -c^2 {\bf P}^2 = c^4 M^2 ,
\end{align}
provides us with an equation, linking $M$, the invariant rest mass, and $E$, the conserved total energy. The total energy equals the Hamiltonian of the system
since we are using canonical coordinates from the very beginning. It is a requirement for the notion of center and total momentum coordinates that $M$ is not depending on external coordinates
\begin{align}
 \{X_i,M\} =&0,    
 &  
 \{P_i,M\} =&0,
\end{align}
applying the canonical commutation relations
\begin{align}
\left\{X_i,\, P_j\right\} = \delta_{ij}\,, \qquad \left\{x_i,\, p_j\right\} = \delta_{ij}\,, \qquad \hbox{zero otherwise}.
\end{align}

The Hamiltonian determines the equation of motion of any function $f$ on phase space,
\begin{align}
 \dot{f} =& \{f,H\} = \frac{1}{2H} \{f,H^2\} = \frac{1}{2H} \Big( c^4\{f,M^2\} +c^2\{f,{\bf P}^2\} \Big),
\end{align}
in particular,
\begin{align}
 \dot{X}^k &= \frac{c^2}{H}  P^k,
 &
 \dot{x}^k &= \frac{c^4}{2H} \{x^k,M^2\},
 \\
 \dot{P}^k &= 0,
 &
 \dot{p}^k &= \frac{c^4}{2H} \{p^k,M^2\}.
\end{align}

Usually, in the binary problem, it is sufficient to just set $P^k =0$ in order to transform to the rest frame, e.g. \cite{DJS00a}. 
In this paper though, we are interested in finding coordinates without resorting to this shortcut. We say that $M$ is decoupling, if it is only depending on internal coordinates.

Let us summarize the properties the new coordinates should have. First, the new set of coordinates has to be a canonical set of coordinates, so that the Hamiltonian form of the equations of motion is preserved. Second, the total linear and angular momenta have to be preserved in their forms and are therefore fixpoints of the transformation. Third, the equations of motion should decouple as described above. Last, the Newtonian limit should reproduce the known Newtonian two-body Hamiltonian in center-of-mass and relative coordinates, which is ensured because we do not leave the post-Newtonian framework. 

The first requirement is realised by introducing a canonical transformation $T_g$, which will raise the Newtonian coordinates to a given higher order post-Newtonian level:
\begin{align}\label{transformation}
\xz & = T_{g}\xz^{\op{N}}\ , &
\xs & = T_{g}\xs^{\op{N}} ,\nonumber \\
\pz & = T_{g}\pz^{\op{N}}\ ,&
\ps & = T_{g}\ps^{\op{N}} .
\end{align}

The transformation $T_g$ can be represented in terms of its infinitesimal generator $g$ by a Lie-series of the form 
\begin{eqnarray}\label{T_g}
T_{g}A = e^{\{.\,,\,g\}}A \equiv A + \left\{A,\, g\right\}+\frac{1}{2}\left\{\left\{A,\, g\right\},\, g\right\}+\frac{1}{6}\left\{\left\{\left\{A,\, g\right\},\, g\right\},\, g\right\}+\dots\, .
\end{eqnarray}
Clearly, $T_g$ written in this way, defines a canonical transformation.

The second requirement then translates into the following conditions:
\begin{align}
\left\{\mathbf{P},\, g\right\} & =  0, &
\left\{\mathbf{J},\, g\right\} & =  0,
\end{align}
where $\mathbf{J}$ denotes the total angular momentum, i.e. $\mathbf{J}= \mathbf{L} + \mathbf{S}$ with $\mathbf{S} = \sum_a \mathbf{s}_a$ [actually in our paper, the spins $\mathbf{s}_a$ of the single objects are put equal to zero]. The generator $g$ has been calculated in \cite{RS10} for various PN approximations. For the convenience of the reader, we recap the generators relevant to us in the Appendix. 

The Poincar\'e or inhomogeneous Lorentz group is the fundamental symmetry group of physical systems living in Minkowski spacetime. 
Isolated systems living in space-asymptotically flat spacetime, i.e.\ asymptotic Minkowski spacetime, also have the 
Poincar\'e group as their symmetry group.
This is the case here. However, our systems are not fully relativistic, but only relativistic to some given PN order. Therefore the Poincar\'e group is not the exact symmetry group but only to a given PN order. In particular, this means that the Poincar\'e algebra is only fulfilled to a given PN order as well.

The Poincar\'e algebra is explicitly given by (see \cite{DJS00} for details)
\begin{eqnarray}
\left\{\pzabs^{i},\, H\right\} &=& \left\{J^{i},\, H\right\}=0,\\
\left\{G^{i},\, H\right\}&=&\pzabs^{i},\quad\left\{\pzabs^{i},\pzabs^{j}\right\}=0,\\
\left\{J^{i},\pzabs^{j}\right\}&=&\epsilon^{ijk}\pzabs^{k},\quad\left\{J^{i},\, J^{j}\right\}=\epsilon^{ijk}J^{k},\\
\left\{J^{i},\, G^{j}\right\}&=&\epsilon^{ijk}G^{k},\quad\left\{G^{i},\, P^{j}\right\}=\clight^{-2}H\delta^{ij},\\
\left\{G^{i},\, G^{j}\right\}&=&-\clight^{-2}\epsilon^{ijk}J^{k},
\end{eqnarray}
with $P^i = P_i, J^i = J_i, G^i = G_i$, and where $H,P^i, J^i,G^i$ denote the total Hamiltonian, total linear momentum, total angular momentum 
including spin, and the center-of-energy position vector, respectively. For completeness, we like to note that the conserved Lorentz boost 
vector $K^i$ is given by $K^i = G^i - t P^i $, where $t$ denotes the asymptotic physical time (in flat spacetime, $t$ would be a physical time parameter 
all-over space belonging to some specific inertial reference system). 

The relativistic canonical center position vector can then be written in the form, e.g.\ \cite{RS10},
\begin{eqnarray} \label{com_coordinate}
\xz & = & \frac{\myvec{G}c^2}{H}+\frac{1}{M\left(H+M\clight^{2}\right)}\left(\myvec{J}-\left(\frac{\myvec{G}c^2}{H}\kreu\pz\right)\right)\kreu\pz\,.
\end{eqnarray}
This choice of $\xz$ ensures that $M$ will only depend on internal variables, i.e. 
\begin{align}
 \frac{\partial M}{\partial P^i} &= \{X^i,M\} =0\, , & \frac{\partial M}{\partial X^i} &= \{P^i,M\} =0.
\end{align}
The Poincar\'e algebra also indicates the following Poisson brackets, which we want to add for completeness,
\begin{eqnarray}
\left\{M,\, H \right\}= 0\,, \quad  
\frac{H}{c^2}\left\{{\bf{X}},\, H \right\}= {\bf{P}}\, .
\end{eqnarray}
In our case of non-spinning objects, the canonical center position vector simplifies to 
\begin{eqnarray} 
\xz & = & \frac{\myvec{G}c^2}{H}+\frac{1}{M\left(H+M\clight^{2}\right)}\left(\myvec{L}-\left(\frac{\myvec{G}c^2}{H}\kreu\pz\right)\right)\kreu\pz\,.
\end{eqnarray}
Notice that the canonical center position vector does not coincide in general with the center-of-energy (or center-of-mass) position vector which is defined by $\frac{\myvec{G}c^2}{H}$, e.g. \cite{GF65}. For a synopsis of the various center definitions see e.g.\ \cite{JS11}.

\section{Canonical center and relative coordinates}

With the aid of the relations in equation (\ref{transformation}) and the generator $g$ derived in \cite{RS10}, we can now calculate the new coordinates explicitly. We use the {\sc Mathematica} \cite{Math} package {\sc XTensor} \cite{xact,xact2}. All files are available upon request.

The coordinate transformations are expressions with many summands, therefore, we split them into terms proportional to $c^{-2n}$.
First, we start with the center coordinate $\xz$, which, 
initially, is identical to the center-of-mass $\mathbf{X}^{\op{N}}$. Second, we present the new relative coordinate $\xs$ and last, the new relative momentum $\ps$.  The total linear momentum is left unchanged, i.e. ${\bf P} = {\bf P}^{\rm N} = {\bf p}_1 + {\bf p}_2$.\\

The new coordinates are functions of $(\mathbf{P}^{\op{N}},\mathbf{x}^{\op{N}},\mathbf{p}^{\op{N}})$ or, after a simple substitution, $(\mathbf{x}^{\op{N}}, 
\mathbf{p}_1,\mathbf{p}_2)$. Apart from $\mathbf{X}$, there is no dependency from $\mathbf{X}^{\op{N}}$ by construction. We use the abbreviation $r_{12}^2 = (\xs^{\op{N}}\cdot\xs^{\op{N}})$.

\subsection{results for $\xz$}
\begin{equation}\label{xBpertubativ}
 \boxed{ X^k  =
  X^{\op{N}k} + X_{\op{1PN}}^k + X_{\op{2PN}}^k + \mathcal{O}(c^{-6})
  }
\end{equation}
	\input{xb_1pn_2pn}

\subsection{results for $\xs$}

\begin{equation}\label{xSpertubativ}
 \boxed{ x^k  =
  x^{\op{N}k} + x_{\op{1PN}}^k + x_{\op{2PN}}^k+\mathcal{O}(c^{-6})
  }
\end{equation}

	\input{xs2pn}

\subsection{results for $\ps$}
\begin{equation}
 \boxed{ p^k  =
  p^{\op{N}k} + p_{\op{1PN}}^k + p_{\op{2PN}}^k+\mathcal{O}(c^{-6})
  }
\end{equation}
	\input{ps2pn}

The total angular momentum of our binary system with non-spinning components, i.e. ${\bf J} = {\bf L} = \sum_a \mathbf{x}_a \times \mathbf{p}_a$,
in center and relative coordinates, reads, 
\begin{equation}
{\bf L}^{\rm N} =  {\bf X}^{\rm N} \times {\bf P}^{\rm N} + {\bf x}^{\rm N} \times {\bf p}^{\rm N} = {\bf X} \times {\bf P} + {\bf x} \times {\bf p} = {\bf L}.
\end{equation}
We confirmed that ${\bf J} = {\bf L}$ is a fixpoint of our canonical transformation.

\section{inverse transformation through 2PN}

It is necessary to know the original Newtonian canonical coordinates in terms of the new center and relative coordinates, in order to substitute the old 
for the new coordinates in a given equation. With a simple further transformation, the standard coordinates for the single components of the binary system 
can then be obtained as functions of the new center and relative coordinates. The inverse transformation through 2PN order reads,
\begin{subequations}\label{xp_inverse}
\begin{align}
  \xs^{\op{N}} &=
  \xs - \xs_{\op{1PN}}- \xs_{\op{2PN}} +\mathcal{O}(c^{-6}), \label{xNinverse}\\
  \ps^{\op{N}} &=
  \ps - \ps_{\op{1PN}}- \ps_{\op{2PN}} +\mathcal{O}(c^{-6}), \label{pNinverse}\\ 
  \mathbf{X}^{\op{N}} &= 
  \mathbf{X} - \mathbf{X}_{\op{1PN}} - \mathbf{X}_{\op{2PN}} + \mathcal{O}(c^{-6}). \label{XNinverse}  
  \end{align}
\end{subequations}
The equations (\ref{xp_inverse}) are simply rearrangements of previous relations and the single terms $\xs_{\op{1PN}},\xs_{\op{2PN}},\ps_{\op{1PN}},\ps_{\op{2PN}},
\mathbf{X}_{\op{1PN}}, \mathbf{X}_{\op{2PN}}$ are still functions of $(\xs^{\op{N}},\myvec{p}_1,\myvec{p}_2)$. Now we are going to invert the transformations step by 
step. In a first step we replace simultaneously $\xs^{\op{N}}$ with $\xs$ and $\ps^{\op{N}}$ with $\ps$ on the right hand side of equations (\ref{xp_inverse}). 
Then we regroup the terms proportional to $c^{-2}$ and find the inverse transformation to first order. The first order inverse transformation is then used in 
another step to derive the second order inverse transformation.
%
	\input{inverseTransform}
%
%

We thus found the old coordinates as functions of the new coordinates. Notice the tremendous simplification of the relations in the rest frame where 
${\bf P} = 0$ holds.

\section{decoupling the rest mass}

In this last section, we want to show that our newly derived coordinates do as promised, i.e.\  $M$ decouples in the new coordinates up to 2PN approximation, analogously to the Newtonian case. For convenience, let us restate equation (\ref{energymomentumequation}):
\begin{equation}
 M^2 =\frac{1}{c^4}H^2 - \frac{1}{c^2}{\bf P}^2 .
\end{equation}
After squaring the Hamiltonian $H= mc^2 + H_{\op{N}} + \frac{1}{c^2}H_{\op{1PN}} + \frac{1}{c^4}H_{\op{2PN}} + \mathcal{O}(c^{-6})$ and grouping all terms together that are proportional to a factor in $c^{-2n}$, we find the following structure:
\begin{align}
 M^2 =\,& E_0 + \frac{1}{c^2} E_1+ \frac{1}{c^4} E_2 + \frac{1}{c^6} E_3 +\mathcal{O}(c^{-8}) .
\end{align}
In the spirit of the perturbative nature of post-Newtonian theory, we expand $M$ with the parameter $c^{-2}$ in a Taylor series:
\begin{align}
 M = \sqrt{E_0}  
  + \frac{1}{c^2} \frac{E_1}{E_0^{\frac{1}{2}}}
  + \frac{1}{c^4} \frac{4 E_2 E_0 -(E_1)^2}{8 E_0^{\frac{3}{2}}}  
  + \frac{1}{c^6} \frac{E_1^3-4 E_1 E_2 E_0 +8 E_3 E_0^2}{16 E_0^{\frac{5}{2}}}  
  + \mathcal{O}(c^{-8})
\end{align}
and find,
\begin{align}\label{kHN}
  M &=
 m_1 +m_2 +\frac{1}{c^2} \Bigg[
  - \frac{G m_1 m_2}{r_{12}}
  + \frac{\left(m_1+m_2\right) (\mathbf{p}^{\op{N}})^2}{2 m_1 m_2}
  \Bigg] \\  
  & \quad
  + \frac{1}{c^4} \Bigg[
    \frac{G^2 m_1 m_2 \left(m_1+m_2\right)}{2 r_{12}^2}
  - \frac{1}{8 m_1^3 m_2^3 \left(m_1+m_2\right)}\Big(
      4 m_1^2 m_2^2 (\mathbf{P}^{\op{N}}\cdot\mathbf{p}^{\op{N}})^2
 \nonumber\\  & \qquad\qquad
    + 4 m_1 m_2 \left(-m_1{}^2+m_2{}^2\right) (\mathbf{P}^{\op{N}}\cdot\mathbf{p}^{\op{N}}) (\mathbf{p}^{\op{N}})^2
    + \left(m_1^4 +m_1^3 m_2 +m_1 m_2^3 +m_2^4\right) (\mathbf{p}^{\op{N}})^4
    \Big)
 \nonumber\\  & \qquad
  - \frac{1}{2 m_1 m_2 \left(m_1+m_2\right)^2 r_{12}^3}G \Big(
      m_1 m_2 \left(m_1^2-m_2^2\right) r_{12}^2 (\mathbf{P}^{\op{N}}\cdot\mathbf{p}^{\op{N}})
 \nonumber\\  & \qquad\qquad
    + \left(m_1+m_2\right)^2 \left(3 m_1^2 +7 m_1 m_2 +3 m_2^2\right) r_{12}^2 (\mathbf{p}^{\op{N}})^2
 \nonumber\\  & \qquad\qquad
    + m_1 m_2 \Big(
      -m_1 m_2 (\mathbf{P}^{\op{N}}\cdot\mathbf{x}^{\op{N}})^2
 \nonumber\\  & \qquad\qquad\qquad
      + (m_1^2 -m_2^2) (\mathbf{P}^{\op{N}}\cdot\mathbf{x}^{\op{N}}) (\mathbf{p}^{\op{N}}\cdot\mathbf{x}^{\op{N}})
      + (m_1 +m_2)^2 (\mathbf{p}^{\op{N}}\cdot\mathbf{x}^{\op{N}})^2
      \Big)
  \Big)
  \Bigg] . \nonumber
\end{align}
It is obvious that $M$ does not decouple in the old Newtonian coordinates. 

We now substitute for the new coordinates in $M$:
\begin{align}
 M&=
  m_1 + m_2
  + \frac{1}{c^2} \Bigg[
      \frac{(m_1+m_2) (\mathbf{p})^2}{2 m_1 m_2}
    - \frac{G m_1 m_2}{\sqrt{(\mathbf{x}\cdot\mathbf{x})}}
  \Bigg]
\nonumber\\ & \quad
  + \frac{1}{c^4} \Bigg[
    - \frac{(m_1^3 +m_2^3) }{8 m_1^3 m_2^3}\,(\mathbf{p})^4
    + \frac{G^2 m_1 m_2 (m_1+m_2)}{2 (\mathbf{x}\cdot\mathbf{x})}
\nonumber\\ & \qquad
    + \frac{G}{2 m_1 m_2 (\mathbf{x}\cdot\mathbf{x})^{3/2}} 
    \Big(
      - m_1 m_2 (\mathbf{p}\cdot\mathbf{x})^2
      - (3 m_1^2 +7 m_1 m_2 +3 m_2^2) (\mathbf{p})^2 (\mathbf{x})^2
    \Big)
  \Bigg]
\nonumber\\ & \quad
  + \frac{1}{c^6} \Bigg[
      \frac{(m_1^5 +m_2^5)}{16 m_1^5 m_2^5} (\mathbf{p})^6
    - \frac{G^3 m_1 m_2 (m_1^2 +5 m_1 m_2 +m_2^2)}{4 (\mathbf{x}\cdot\mathbf{x})^{3/2}}
\nonumber\\ & \qquad
    + G^2 \frac{ (m_1+m_2)}{2 m_1 m_2 (\mathbf{x}\cdot\mathbf{x})^2} \Big(
	3 m_1 m_2 (\mathbf{p}\cdot\mathbf{x})^2
      + (5 m_1^2 +18 m_1 m_2 +5 m_2^2) (\mathbf{p})^2 (\mathbf{x})^2
      \Big)
\nonumber\\ & \qquad
    + \frac{G }{8 m_1^3 m_2^3 (\mathbf{x}\cdot\mathbf{x})^{5/2}}
    \Big(
      - 3 m_1^2 m_2^2 (\mathbf{p}\cdot\mathbf{x})^4-2 m_1^2 m_2^2 (\mathbf{p})^2 (\mathbf{p}\cdot\mathbf{x})^2 (\mathbf{x}\cdot\mathbf{x})
\nonumber\\ & \qquad\qquad
      + (5 m_1^4 -13 m_1^2 m_2^2 +5 m_2^4) (\mathbf{p})^4 (\mathbf{x}\cdot\mathbf{x})^2
    \Big)
  \Bigg] .
\label{kHnew}
\end{align}
We see, very nicely, that $M$, in our newly derived coordinates, solely depends on relative variables and thus decouples. 


\acknowledgments 
We like to thank J. Hartung and J. Steinhoff for developing PNtools, a Mathematica package, that we used.
IG gratefully acknowledges support by the University of Cape Town Faculty of Science PhD Fellowship Program.

\bibliographystyle{amsalpha}
\bibliography{citations}

\begin{appendix}
\section{PN generators}
Rothe and Sch\"afer derived the generator $g$ for the obtention of the canonical center and relative coordinates in \cite{RS10}, denoted therein
$g_{\rm point}$ (``point'' meaning without spin). Also, the abbreviation ${\bf n}_{12} = ({\bf x}_1 - {\bf x}_2)/r_{12}$ is used. 

	\input{generat2pn}

\end{appendix}

\end{document}

%% file: xb_1pn_2pn.tex
\begin{align*}
  X_{\op{1PN}}^k &=
  \frac{1}{c^2} 
  \Bigg[ (x_1^k - x_2^k)
    \Big( -\frac{m_1^2 ( \mathbf{p}_2)^2 }{2 m_2 m^3}
      +\frac{ m_2^2 (\mathbf{p}_1)^2}{2 m_1 m^3}
      +\frac{(m_1- m_2)( \mathbf{p}_1 \cdot  \mathbf{p}_2)}{2 m^3}
      +\frac{G  m_1 m_2(m_1- m_2)}{2 r_{12}m^2}
    \Big)
\\[0.2cm] & \qquad
    + \frac{(m_2\, p_1^k - m_1\, p_2^k )}{2m^3}
    \Big( 
	(\mathbf{p}_1+ \mathbf{p}_2) \cdot (\mathbf{x}_1 - \mathbf{x}_2)
    \Big)
  \Bigg]
 \label{X1pn}
\end{align*}

 {\allowdisplaybreaks
\begin{align}
 X_{\op{2PN}}^k &=
  \frac{1}{c^4} \Bigg[
  p_1^k \Bigg(
  (\mathbf{p}_1\cdot(\mathbf{x}_1-\mathbf{x}_2))\Big(
    - \frac{G m_1 (m_1-5 m_2) m_2}{4 (m_1+m_2)^4 r_{12}}
\nonumber\\ & \qquad\qquad\qquad
    - \frac{3 m_2 (m_1+2 m_2) (\mathbf{p}_1)^2}{8 m_1 (m_1+m_2)^5}
    + \frac{3 m_2 (\mathbf{p}_1\cdot\mathbf{p}_2)}{4 (m_1+m_2)^5}
    - \frac{(-2 m_1^2 +2 m_1 m_2 +m_2^2) (\mathbf{p}_2)^2}{8 m_2 (m_1+m_2)^5}
  \Big) 
\nonumber\\ & \qquad\qquad
  + (\mathbf{p}_2\cdot(\mathbf{x}_1-\mathbf{x}_2))\ \Big(
    - \frac{G (5 m_1^3 +16 m_1^2 m_2 +10 m_1 m_2^2 +5 m_2^3)}{4 (m_1+m_2)^4 r_{12}}
\nonumber\\ & \qquad\qquad\qquad
    - \frac{3 m_2 (m_1+2 m_2)(\mathbf{p}_1)^2}{8 m_1 (m_1+m_2)^5}
    + \frac{3 m_2 (\mathbf{p}_1\cdot\mathbf{p}_2)}{4 (m_1+m_2)^5}
    - \frac{(-2 m_1^2 +2 m_1 m_2 +m_2^2) (\mathbf{p}_2)^2}{8 m_2 (m_1+m_2)^5}
  \Big) 
  \Bigg)
\nonumber\\ & \qquad
  + p_2^k\ \Bigg(
   (\mathbf{p}_1\cdot(\mathbf{x}_1-\mathbf{x}_2))\Big(
      \frac{G (5 m_1^3 +10 m_1^2 m_2 +16 m_1 m_2^2 +5 m_2^3)}{4 (m_1+m_2)^4 r_{12}}
\nonumber\\ & \qquad\qquad\qquad
    + \frac{(m_1^2 +2 m_1 m_2 -2 m_2^2) (\mathbf{p}_1)^2}{8 m_1 (m_1+m_2)^5}
    - \frac{3 m_1 (\mathbf{p}_1\cdot\mathbf{p}_2)}{4 (m_1+m_2)^5}
    + \frac{3 m_1 (2 m_1+m_2) (\mathbf{p}_2)^2}{8 m_2 (m_1+m_2)^5}
  \Big)
\nonumber\\ & \qquad\qquad
  + (\mathbf{p}_2\cdot(\mathbf{x}_1-\mathbf{x}_2))\ \Big(
      \frac{G m_1 m_2 (-5 m_1+m_2)}{4 (m_1+m_2)^4 r_{12}}
\nonumber\\ & \qquad\qquad\qquad
    + \frac{(m_1^2 +2 m_1 m_2-2 m_2^2) (\mathbf{p}_1)^2}{8 m_1 (m_1+m_2)^5}
    - \frac{3 m_1 (\mathbf{p}_1\cdot\mathbf{p}_2)}{4 (m_1+m_2)^5}
    + \frac{3 m_1 (2 m_1+m_2) (\mathbf{p}_2)^2}{8 m_2 (m_1+m_2)^5}
  \Big) 
  \Bigg)
\nonumber\\ & \qquad
  + (x_1^k- x_2^k)\Bigg(
    - \frac{G^2 m_1 (m_1-m_2) m_2 (m_1^2 +m_2^2)}{4 (m_1+m_2)^3 r_{12}^2}
    - \frac{m_2^2 (m_1^2 +5 m_1 m_2 +m_2^2) }{8 m_1^3 (m_1+m_2)^5}\ (\mathbf{p}_1)^4
\nonumber\\ & \qquad\qquad
    + \frac{3 (m_1-m_2) }{4 (m_1+m_2)^5}\ (\mathbf{p}_1\cdot\mathbf{p}_2)^2
    + \frac{G m_1 (5 m_1^3 +6 m_1^2 m_2 +6 m_1 m_2^2 -m_2^3) }{4 m_2 (m_1+m_2)^4 r_{12}}\ (\mathbf{p}_2)^2
\nonumber\\ & \qquad\qquad
    + \frac{m_1^2 (m_1^2 +5 m_1 m_2 +m_2^2) }{8 m_2^3 (m_1+m_2)^5}\ (\mathbf{p}_2)^4
    + \frac{G m_2 (m_1^3 -6 m_1^2 m_2 -6 m_1 m_2^2 -5 m_2^3)}{4 m_1 (m_1+m_2)^4 r_{12}}\ (\mathbf{p}_1)^2 
\nonumber\\ & \qquad\qquad
    - \frac{(m_1^2 +5 m_1 m_2 -8 m_2^2)}{8 m_1 (m_1+m_2)^5}\ (\mathbf{p}_1)^2 (\mathbf{p}_1\cdot\mathbf{p}_2)
    + \frac{3 (m_1-m_2) }{8 (m_1+m_2)^5}\ (\mathbf{p}_1)^2(\mathbf{p}_2)^2
\nonumber\\ & \qquad\qquad
    - \frac{G (m_1-m_2) (7 m_1^2 +8 m_1 m_2 +7 m_2^2)}{4 (m_1+m_2)^4 r_{12}}\ (\mathbf{p}_1\cdot\mathbf{p}_2)
\nonumber\\ & \qquad\qquad
    + \frac{(-8 m_1^2 +5 m_1 m_2 +m_2^2) }{8 m_2 (m_1+m_2)^5}\ (\mathbf{p}_2)^2(\mathbf{p}_1\cdot\mathbf{p}_2)
\nonumber\\ & \qquad\qquad
    - \frac{G (m_1-m_2)  }{4 (m_1+m_2)^2 r_{12}^3}\Big(
      (\mathbf{p}_1\cdot (\mathbf{x}_1-\mathbf{x}_2)) (\mathbf{p}_2\cdot (\mathbf{x}_1-\mathbf{x}_2))
    \Big)
  \Bigg)
  \Bigg]
\end{align}
}

%% file: xs2pn.tex
\begin{align}
  x_{\op{1PN}}^k &=
  \frac{1}{c^2}\left(
    \frac{(-m_1 +2 m_2) }{ 2 m_1 (m_1+m_2)^2} p_1^k
  + \frac{(2 m_1 -m_2) }{ 2 m_2 (m_1+m_2)^2} p_2^k
  \right) 
  \Big( (\mathbf{p}_1+\mathbf{p}_2)\cdot \mathbf{x}^{\op{N}} \Big)
\end{align}



\begin{align}
  x_{\op{2PN}}^k&=
  \frac{1}{c^4} \Bigg[
  p_1^k \Bigg(
    \Bigg( \frac{G (7 m_1^3 -4 m_1^2 m_2 -9 m_1 m_2^2- 10 m_2^3)}{4 m_1 (m_1 +m_2)^3 r_{12}}
  \\ &\qquad\qquad\qquad
      - \frac{(-3 m_1^3 +4 m_1^2 m_2 + 8 m_1 m_2^2 +4 m_2^3) (\mathbf{p}_1)^2}{8 m_1^3 (m_1 +m_2)^4}
 \nonumber \\ &\qquad\qquad\qquad
      + \frac{(-2 m_1^3 -m_1^2 m_2 +2 m_1 m_2^2 +4 m_2^3) (\mathbf{p}_1\cdot\mathbf{p}_2)}{4 m_1^2 m_2 (m_1 +m_2)^4}
 \nonumber \\ &\qquad\qquad\qquad
      + \frac{(4 m_1^2 +3 m_1 m_2 -4 m_2^2) (\mathbf{p}_2)^2}{8 m_1 m_2 (m_1+m_2)^4}
    \Bigg) \Big((\mathbf{p}_1+\mathbf{p}_2)\cdot \mathbf{x}^{\op{N}} \Big)
    \Bigg)
 \nonumber \\ &\qquad
  + p_2^k \Bigg(
    \Bigg( -\frac{G (10 m_1^3 +9 m_1^2 m_2 +4 m_1 m_2^2 -7 m_2^3)}{4 m_2 (m_1 +m_2)^3 r_{12}}
 \nonumber \\ &\qquad\qquad\qquad
      + \frac{(-4 m_1^2 +3 m_1 m_2 +4 m_2^2) (\mathbf{p}_1)^2}{8 m_1 m_2 (m_1 +m_2)^4}
 \nonumber \\ &\qquad\qquad\qquad
      + \frac{(4 m_1^3 +2 m_1^2 m_2 -m_1 m_2^2 -2 m_2^3) (\mathbf{p}_1\cdot\mathbf{p}_2)}{4 m_1 m_2^2 (m_1+m_2)^4}
 \nonumber \\ &\qquad\qquad\qquad
      - \frac{(4 m_1^3 +8 m_1^2 m_2 +4 m_1 m_2^2 -3 m_2^3) (\mathbf{p}_2)^2}{8 m_2^3 (m_1+m_2)^4}
    \Bigg) \Big( (\mathbf{p}_1+\mathbf{p}_2)\cdot \mathbf{x}^{\op{N}} \Big)
  \Bigg) 
 \nonumber \\ &\qquad
  + x^{\op{N}k}\Bigg(
    \frac{5 G }{8 (m_1+m_2)r_{12}}(\mathbf{p}_1+\mathbf{p}_2)^2
 \nonumber \\ &\qquad\qquad\quad
    - \frac{G  (m_1 -m_2) }{8 (m_1 +m_2)^3 r_{12}^3}
    \Big((m_1 -5 m_2)(\mathbf{p}_1\cdot\mathbf{x}^{\op{N}}) 
 \nonumber \\ &\hspace{5cm}
      + (5m_1 - m_2)(\mathbf{p}_2\cdot\mathbf{x}^{\op{N}})
      \Big) \Big( (\mathbf{p}_1+\mathbf{p}_2)\cdot \mathbf{x}^{\op{N}} \Big)
    \Bigg)
  \Bigg]
 \nonumber
\end{align}

%% file: ps2pn.tex
\begin{align}
  p_{\op{1PN}}^k&=
  \frac{1}{c^2} \Bigg[
  (p_1^k+p_2^k) \Bigg(
    \frac{G m_1 m_2 (-m_1+m_2)}{2 (m_1+m_2)^2 r_{12}}
    - \frac{m_2^2 (\mathbf{p}_1)^2}{2 m_1 (m_1+m_2)^3}
 \nonumber\\  & \qquad\qquad\quad
    + \frac{(-m_1+m_2) (\mathbf{p}_1\cdot\mathbf{p}_2)}{2(m_1+m_2)^3}
    + \frac{m_1^2 (\mathbf{p}_2)^2}{2 m_2 (m_1+m_2)^3}
    \Bigg)
 \nonumber\\  & \qquad\quad
  + x^{\op{N}k}\ \frac{G m_1 m_2 (m_1-m_2) }{2 (m_1+m_2)^2 r_{12}^3} \big((\mathbf{p}_1+\mathbf{p}_2)\cdot\mathbf{x}^{\op{N}}\big)
  \Bigg]
\end{align}

\clearpage
{\allowdisplaybreaks
\begin{align}
p_{\op{2PN}}^k &=
  \frac{1}{c^4} \Bigg[
  (p_1^k+p_2^k) \Bigg(
    \frac{G^2 m_1 (m_1-m_2) m_2 (m_1^2+m_2^2)}{4 (m_1+m_2)^3 r_{12}^2}
    + \frac{m_2^2 (m_1^2 +5 m_1 m_2 +m_2^2) (\mathbf{p}_1)^4}{8 m_1^3 (m_1+m_2)^5}
\nonumber\\ &\qquad\qquad\quad
    - \frac{3 (m_1-m_2) (\mathbf{p}_1\cdot\mathbf{p}_2)^2}{4 (m_1 +m_2)^5}
    - \frac{m_1^2 (m_1^2 +5 m_1 m_2 +m_2^2) (\mathbf{p}_2)^4}{8 m_2^3 (m_1+m_2)^5}
\nonumber\\ &\qquad\qquad\quad
    + (\mathbf{p}_1)^2 \Big(
      \frac{(m_1^2 +5m_1 m_2 -8 m_2^2) (\mathbf{p}_1\cdot\mathbf{p}_2)}{8 m_1 (m_1+m_2)^5}
      - \frac{3 (m_1-m_2) (\mathbf{p}_2)^2}{8 (m_1+m_2)^5}
      \Big)
\nonumber\\ &\qquad\qquad\quad
    + (\mathbf{p}_1\cdot\mathbf{p}_2) \left(
      - \frac{(-8 m_1^2 +5 m_1 m_2 +m_2^2)(\mathbf{p}_2)^2}{8 m_2 (m_1+m_2)^5}
      \right)
  \Bigg)
\nonumber\\ &\qquad
  + p_1^k \Bigg(
    - \frac{G (10 m_1^4 +12 m_1^3 m_2 +17 m_1^2 m_2^2 +8 m_1 m_2^3 +5 m_2^4) (\mathbf{p}_2)^2}{8 m_2 (m_1+m_2)^4 r_{12}}
\nonumber\\ &\qquad\qquad\quad
    + \frac{G m_2 (-7 m_1^3 +2 m_1^2 m_2 +7 m_1 m_2^2 +10 m_2^3)}{8 m_1 (m_1 +m_2)^4 r_{12}}(\mathbf{p}_1)^2 
\nonumber\\ &\qquad\qquad\quad
    + \frac{G (7 m_1^3 -4 m_1^2 m_2 -11 m_1 m_2^2 -12 m_2^3)}{4 (m_1+m_2)^4 r_{12}}(\mathbf{p}_1\cdot\mathbf{p}_2)
\nonumber\\ &\qquad\qquad\quad
    + \frac{G m_2 (-m_1+m_2) (\mathbf{p}_1\cdot\mathbf{x}^{\op{N}})^2}{8 (m_1+m_2)^3 r_{12}^3}
    + \frac{G (m_1-m_2) (m_1+2 m_2) (\mathbf{p}_1\cdot\mathbf{x}^{\op{N}}) (\mathbf{p}_2\cdot\mathbf{x}^{\op{N}})}{4 (m_1+m_2)^3 r_{12}^3}
\nonumber\\ &\qquad\qquad\quad
    + \frac{3 G (m_1-m_2) m_2 (\mathbf{p}_2\cdot\mathbf{x}^{\op{N}})^2}{8 (m_1+m_2)^3 r_{12}^3}
    \Bigg)
\nonumber\\ &\qquad
  + p_2^k \Bigg(
    - \frac{G m_1 (10 m_1^3 +7 m_1^2 m_2 +2 m_1 m_2^2 -7 m_2^3) (\mathbf{p}_2)^2}{8 m_2 (m_1+m_2)^4 r_{12}}
\nonumber\\ &\qquad\qquad\quad
    + \frac{G (5 m_1^4+8 m_1^3 m_2 +17 m_1^2 m_2^2 +12 m_1 m_2^3 +10 m_2^4)}{8 m_1 (m_1+m_2)^4 r_{12}} (\mathbf{p}_1)^2
\nonumber\\ &\qquad\qquad\quad
    + \frac{G (12 m_1^3 +11 m_1^2 m_2 +4 m_1 m_2^2 -7 m_2^3)}{4 (m_1+m_2)^4 r_{12}}(\mathbf{p}_1\cdot\mathbf{p}_2)
\nonumber\\ &\qquad\qquad\quad
    + \frac{3 G m_1 (m_1-m_2) (\mathbf{p}_1\cdot\mathbf{x}^{\op{N}})^2}{8 (m_1+m_2)^3 r_{12}^3}
    + \frac{G (m_1-m_2) (2 m_1+m_2) (\mathbf{p}_1\cdot\mathbf{x}^{\op{N}}) (\mathbf{p}_2\cdot\mathbf{x}^{\op{N}})}{4 (m_1+m_2)^3 r_{12}^3}
\nonumber\\ &\qquad\qquad\quad
    + \frac{G m_1 (-m_1+m_2) (\mathbf{p}_2\cdot\mathbf{x}^{\op{N}})^2}{8 (m_1+m_2)^3 r_{12}^3}
    \Bigg)
\nonumber\\ &\qquad
  + x^{\op{N}k}\Bigg(
    \frac{3 G (m_1-m_2)^2 m_2 (\mathbf{p}_1\cdot\mathbf{x}^{\op{N}})^3}{8 (m_1+m_2)^4 r_{12}^5}
    - \frac{G^2 m_1 (m_1-m_2) m_2 (m_1^2 +m_2^2) (\mathbf{p}_2\cdot\mathbf{x}^{\op{N}})}{2 (m_1+m_2)^3 r_{12}^4}
\nonumber\\ &\qquad\qquad\quad
    - \frac{3 G (m_1-m_2) (3 m_1^2 +m_1 m_2 +4 m_2^2) (\mathbf{p}_1\cdot\mathbf{x}^{\op{N}})^2 (\mathbf{p}_2\cdot\mathbf{x}^{\op{N}})}%
	{8 (m_1+m_2)^4 r_{12}^5}
\nonumber\\ &\qquad\qquad\quad
    - \frac{3 G m_1 (m_1-m_2)^2 (\mathbf{p}_2\cdot\mathbf{x}^{\op{N}})^3}{8 (m_1+m_2)^4 r_{12}^5}
\nonumber\\ &\qquad\qquad\quad
    + (\mathbf{p}_1\cdot\mathbf{p}_2) \left(
      - \frac{5 G (m_1-2 m_2) (\mathbf{p}_1\cdot\mathbf{x}^{\op{N}})}{4 (m_1+m_2)^2 r_{12}^3}
      + \frac{5 G (-2 m_1+m_2) (\mathbf{p}_2\cdot\mathbf{x}^{\op{N}})}{4 (m_1+m_2)^2 r_{12}^3}
      \right)
\nonumber\\ &\qquad\qquad\quad
    + (\mathbf{p}_2)^2 \Big(
      \frac{G (10 m_1^3 +6 m_1^2 m_2 +5 m_1 m_2^2 +5 m_2^3) (\mathbf{p}_1\cdot\mathbf{x}^{\op{N}})}{8 m_2 (m_1+m_2)^3 r_{12}^3}
\nonumber\\ &\qquad\qquad\qquad
      + \frac{G m_1 (10 m_1^2 +m_1 m_2 -5 m_2^2) (\mathbf{p}_2\cdot\mathbf{x}^{\op{N}})}{8 m_2 (m_1+m_2)^3 r_{12}^3}
      \Big)
\nonumber\\ &\qquad\qquad\quad
    + (\mathbf{p}_1)^2 \Big(
      - \frac{G m_2 (-5 m_1^2 +m_1 m_2 +10 m_2^2) (\mathbf{p}_1\cdot\mathbf{x}^{\op{N}})}{8 m_1 (m_1+m_2)^3 r_{12}^3}
\nonumber\\ &\qquad\qquad\qquad
      - \frac{G (5 m_1^3 +5 m_1^2 m_2 +6 m_1 m_2^2 +10 m_2^3) (\mathbf{p}_2\cdot\mathbf{x}^{\op{N}})}{8 m_1 (m_1+m_2)^3 r_{12}^3}
      \Big)
\nonumber\\ &\qquad\qquad\quad
    + (\mathbf{p}_1\cdot\mathbf{x}^{\op{N}}) \Big(
      - \frac{G^2 m_1 (m_1-m_2) m_2 (m_1^2 +m_2^2)}{2 (m_1 +m_2)^3 r_{12}^4}
\nonumber\\ &\qquad\qquad\qquad
      - \frac{3 G (m_1-m_2) (4 m_1^2 +m_1 m_2 +3 m_2^2) (\mathbf{p}_2\cdot\mathbf{x}^{\op{N}})^2}{8 (m_1+m_2)^4 r_{12}^5}
      \Big)
    \Bigg)
  \Bigg]
\end{align}

}

%% file: inverseTransform.tex
\begin{align}
  x^{\text{N}k} &=
  x^k
  \nonumber \\ &
  + \frac{1}{c^2} \Bigg[
    - P^k \frac{(\mathbf{P}\cdot\mathbf{x})}{2 (m_1+m_2)^2}
    + p^k \frac{(m_1-m_2) (\mathbf{P}\cdot\mathbf{x})}{m_1^2 m_2+m_1 m_2^2}
    \Bigg]
  \nonumber \\ &
  + \frac{1}{c^4} \Bigg[
    P^k \Bigg(
      \frac{3}{8 (m_1+m_2)^4} (\mathbf{P})^2 (\mathbf{P}\cdot\mathbf{x})
    + \frac{(-m_1+m_2)}{2 m_1 m_2 (m_1+m_2)^3} (\mathbf{P}\cdot\mathbf{p})(\mathbf{P}\cdot\mathbf{x})
  \nonumber \\ & \quad\quad
    + \frac{1}{2 m_1 m_2 (m_1+m_2)^2} (\mathbf{p})^2(\mathbf{P}\cdot\mathbf{x})
    + G \frac{ (5 m_1^2+6 m_1 m_2+5 m_2^2)}{4 (m_1+m_2)^3 \sqrt{(\mathbf{x}\cdot\mathbf{x})}} (\mathbf{P}\cdot\mathbf{x})
    \Bigg)
  \nonumber \\ & \quad
    + p^k \Bigg(
    \frac{(-m_1+m_2) (\mathbf{P})^2 (\mathbf{P}\cdot\mathbf{x})}{2 m_1 m_2 (m_1+m_2)^3}
    + \frac{(m_1^2-m_1 m_2+m_2^2) }{m_1^2 m_2^2 (m_1+m_2)^2} (\mathbf{P}\cdot\mathbf{p})(\mathbf{P}\cdot\mathbf{x}) 
  \nonumber \\ & \quad\quad
    + \frac{(-m_1^2+m_2^2)}{2 m_1^3 m_2^3}  (\mathbf{p})^2(\mathbf{P}\cdot\mathbf{x}) 
    + G\ \frac{(-5m_1^3-3m_1^2m_2+3m_1m_2^2+5m_2^3)}{2m_1m_2(m_1+m_2)^2 \sqrt{(\mathbf{x}\cdot\mathbf{x})}} 
    (\mathbf{P}\cdot\mathbf{x}) 
    \Bigg)
  \nonumber \\ & \quad
    + x^k \Bigg(
      - G \frac{3 (m_1-m_2)^2 (\mathbf{P}\cdot\mathbf{x})^2}{8 (m_1+m_2)^3 (\mathbf{x}\cdot\mathbf{x})^{3/2}}
      + G \frac{(-m_1+m_2) (\mathbf{P}\cdot\mathbf{x}) (\mathbf{p}\cdot\mathbf{x})}{2 (m_1+m_2)^2 (\mathbf{x}\cdot\mathbf{x})^{3/2}}
  \nonumber \\ & \quad\quad
      - G \frac{5 (\mathbf{P})^2}{8 (m_1+m_2) \sqrt{(\mathbf{x}\cdot\mathbf{x})}}
    \Bigg)
  \Bigg]
\end{align}
\begin{align}
  p^{\text{N}k} &=
  p^k
  \nonumber \\ &
  + \frac{1}{c^2} \Bigg[
  P^k \Bigg(
    \frac{(\mathbf{P}\cdot\mathbf{p})}{2 (m_1+m_2)^2}
    + \frac{(-m_1+m_2) (\mathbf{p})^2}{2 m_1 m_2 (m_1+m_2)}
    + \frac{G m_1 (m_1-m_2) m_2}{2 (m_1+m_2)^2 \sqrt{(\mathbf{x}\cdot\mathbf{x})}}
    \Bigg)
  \nonumber \\ & \quad
  + x^k\ \frac{G m_1 m_2 (-m_1+m_2) (\mathbf{P}\cdot\mathbf{x})}{2 (m_1+m_2)^2 (\mathbf{x}\cdot\mathbf{x})^{3/2}}
  \Bigg]
  \nonumber \\ &
  + \frac{1}{c^4} \Bigg[
  P^k \Bigg(
    - \frac{(\mathbf{P})^2 (\mathbf{P}\cdot\mathbf{p})}{8 (m_1+m_2)^4}
    + \frac{(m_1-m_2) (m_1+m_2) (\mathbf{p})^4}{8 m_1^3 m_2^3}
    -\frac{(\mathbf{p})^2 (\mathbf{P}\cdot\mathbf{p})}{2 m_1 m_2 (m_1+m_2)^2}
  \nonumber \\ & \quad\quad
    -\frac{G (5 m_1^2+6 m_1 m_2+5 m_2^2)}{4 (m_1+m_2)^3 \sqrt{(\mathbf{x}\cdot\mathbf{x})}}(\mathbf{P}\cdot\mathbf{p})
    + \frac{G m_1 m_2 (-m_1+m_2) (\mathbf{P}\cdot\mathbf{x})^2}{4 (m_1+m_2)^4(\mathbf{x}\cdot\mathbf{x})^{3/2}}
  \nonumber \\ & \quad\quad
    + \frac{G (m_1-m_2)^2 (\mathbf{P}\cdot\mathbf{x}) (\mathbf{p}\cdot\mathbf{x})}{4 (m_1+m_2)^3 (\mathbf{x}\cdot\mathbf{x})^{3/2}}
    + \frac{G (m_1-m_2) (\mathbf{p}\cdot\mathbf{x})^2}{4 (m_1+m_2)^2 (\mathbf{x}\cdot\mathbf{x})^{3/2}} 
  \nonumber \\ & \quad\quad
    - \frac{G^2 m_1 (m_1-m_2) m_2 (m_1^2+m_2^2)}{4 (m_1+m_2)^3 (\mathbf{x}\cdot\mathbf{x})}
    + \frac{G (m_1-m_2) (5 m_1^2+8 m_1 m_2+5 m_2^2) (\mathbf{p})^2}{4 m_1 m_2 (m_1+m_2)^2 \sqrt{(\mathbf{x}\cdot\mathbf{x})}}
    \Bigg)
  \nonumber \\ & \quad
  + p^k \Bigg(
    - \frac{G (m_1-m_2)^2 (\mathbf{P}\cdot\mathbf{x})^2}{8 (m_1+m_2)^3(\mathbf{x}\cdot\mathbf{x})^{3/2}}
    + \frac{G (m_1-m_2)   (\mathbf{P}\cdot\mathbf{x}) (\mathbf{p}\cdot\mathbf{x})}{2 (m_1+m_2)^2(\mathbf{x}\cdot\mathbf{x})^{3/2}}
    + \frac{5 G (\mathbf{P})^2}{8 (m_1+m_2) \sqrt{(\mathbf{x}\cdot\mathbf{x})}}
    \Bigg)
  \nonumber \\ & \quad
  + x^k \Bigg(
    \frac{3 G (m_1-m_2)^2 (\mathbf{P}\cdot\mathbf{x})^2 (\mathbf{p}\cdot\mathbf{x})}{8 (m_1+m_2)^3 (\mathbf{x}\cdot\mathbf{x})^{5/2}}
      + \frac{3 G (-m_1+m_2) (\mathbf{P}\cdot\mathbf{x}) (\mathbf{p}\cdot\mathbf{x})^2}{4 (m_1+m_2)^2 (\mathbf{x}\cdot\mathbf{x})^{5/2}}
  \nonumber \\ & \quad\quad
      + \frac{G^2 m_1 (m_1-m_2) m_2 (m_1^2+m_2^2)}{2 (m_1+m_2)^3 (\mathbf{x}\cdot\mathbf{x})^2} (\mathbf{P}\cdot\mathbf{x})
      + \frac{G (3 m_1+m_2) (m_1+3 m_2)}{4 (m_1+m_2)^3 (\mathbf{x}\cdot\mathbf{x})^{3/2}} (\mathbf{P}\cdot\mathbf{p}) (\mathbf{P}\cdot\mathbf{x})
  \nonumber \\ & \quad\quad
    -\frac{G (m_1-m_2) (5 m_1^2+8 m_1 m_2+5 m_2^2)}{4 m_1 m_2 (m_1+m_2)^2 (\mathbf{x}\cdot\mathbf{x})^{3/2}} (\mathbf{p})^2 (\mathbf{P}\cdot\mathbf{x})
  \nonumber \\ & \quad\quad
    + \frac{G m_1 (m_1-m_2) m_2 }{4 (m_1+m_2)^4(\mathbf{x}\cdot\mathbf{x})^{3/2}} (\mathbf{P}\cdot\mathbf{x})(\mathbf{P})^2
    - \frac{5 G }{8 (m_1+m_2)(\mathbf{x}\cdot\mathbf{x})^{3/2}} (\mathbf{p}\cdot\mathbf{x})(\mathbf{P})^2
  \Bigg)
  \Bigg]
\end{align}
\begin{align}
  \frac{1}{r_{12}} &=
  \frac{1}{\sqrt{(\mathbf{x}\cdot\mathbf{x})}}
  \nonumber \\ &
  + \frac{1}{c^2} \Bigg[
    \frac{(\mathbf{P}\cdot\mathbf{x})^2}{2 (m_1+m_2)^2 (\mathbf{x}\cdot\mathbf{x})^{3/2}}
    + \frac{(-m_1+m_2) (\mathbf{P}\cdot\mathbf{x}) (\mathbf{p}\cdot\mathbf{x})}{m_1 m_2 (m_1+m_2) (\mathbf{x}\cdot\mathbf{x})^{3/2}}
    \Bigg]
  \nonumber \\ &
  + \frac{1}{c^4} \Bigg[
    \frac{3(\mathbf{P}\cdot\mathbf{x})^4}{8 (m_1+m_2)^4(\mathbf{x}\cdot\mathbf{x})^{5/2}}
    - \frac{3 (m_1-m_2) (\mathbf{P}\cdot\mathbf{x})^3 (\mathbf{p}\cdot\mathbf{x})}{2 m_1 m_2 (m_1+m_2)^3(\mathbf{x}\cdot\mathbf{x})^{5/2}}
    + \frac{3 (m_1-m_2)^2 (\mathbf{P}\cdot\mathbf{x})^2 (\mathbf{p}\cdot\mathbf{x})^2}{2 m_1^2 m_2^2 (m_1+m_2)^2(\mathbf{x}\cdot\mathbf{x})^{5/2}}
  \nonumber \\ & \quad
    -\frac{G (7 m_1^2+18 m_1 m_2+7 m_2^2) (\mathbf{P}\cdot\mathbf{x})^2}{8 (m_1+m_2)^3(\mathbf{x}\cdot\mathbf{x})^2}
    + \frac{G (m_1-m_2) (5 m_1^2+9 m_1 m_2+5 m_2^2) (\mathbf{P}\cdot\mathbf{x}) (\mathbf{p}\cdot\mathbf{x})}{2 m_1 m_2 (m_1+m_2)^2(\mathbf{x}\cdot\mathbf{x})^2}
  \nonumber \\ & \quad
    -\frac{(\mathbf{P})^2 (\mathbf{P}\cdot\mathbf{x})^2}{2 (m_1+m_2)^4(\mathbf{x}\cdot\mathbf{x})^{3/2}}
    + \frac{(m_1-m_2) (\mathbf{P}\cdot\mathbf{p})(\mathbf{P}\cdot\mathbf{x})^2}{m_1 m_2 (m_1+m_2)^3(\mathbf{x}\cdot\mathbf{x})^{3/2}}
    - \frac{(m_1^3+m_2^3) (\mathbf{p})^2 (\mathbf{P}\cdot\mathbf{x})^2}{2 m_1^2 m_2^2 (m_1+m_2)^3(\mathbf{x}\cdot\mathbf{x})^{3/2}}
  \nonumber \\ & \quad
    + \frac{(m_1-m_2) (\mathbf{P})^2(\mathbf{P}\cdot\mathbf{x}) (\mathbf{p}\cdot\mathbf{x})}{2 m_1 m_2 (m_1+m_2)^3(\mathbf{x}\cdot\mathbf{x})^{3/2}}
    - \frac{(m_1^2-m_1 m_2+m_2^2) (\mathbf{P}\cdot\mathbf{p})}{m_1^2 m_2^2 (m_1+m_2)^2(\mathbf{x}\cdot\mathbf{x})^{3/2}}(\mathbf{P}\cdot\mathbf{x}) (\mathbf{p}\cdot\mathbf{x})
  \nonumber \\ & \quad
    + \frac{(m_1-m_2) (m_1+m_2) (\mathbf{p})^2}{2 m_1^3 m_2^3(\mathbf{x}\cdot\mathbf{x})^{3/2}}(\mathbf{P}\cdot\mathbf{x}) (\mathbf{p}\cdot\mathbf{x})
    + \frac{5 G (\mathbf{P})^2}{8 (m_1+m_2) (\mathbf{x}\cdot\mathbf{x})}
    \Bigg]
\end{align}
\begin{align}
  X^{\text{N}k} &=
  X^k 
  \nonumber \\ &
  + \frac{1}{c^2} \Bigg[
    - p^k\frac{ (\mathbf{P}\cdot\mathbf{x})}{2 (m_1+m_2)^2}
  \nonumber \\ & \quad
    + \Bigg(
      - \frac{(\mathbf{P}\cdot\mathbf{p})}{2 (m_1+m_2)^2}
      + \frac{(m_1-m_2) (\mathbf{p})^2}{2 m_1^2 m_2+2 m_1 m_2^2}
      + \frac{G m_1 m_2 (-m_1+m_2)}{2 (m_1+m_2)^2 \sqrt{(\mathbf{x}\cdot\mathbf{x})}}\Bigg)
    x^k\Bigg]
  \nonumber \\ &
  + \frac{1}{c^4} \Bigg[
    p^k\Bigg(
      \Big( 
	\frac{3 (\mathbf{P})^2}{8 (m_1+m_2)^4}
	+ \frac{(-m_1+m_2) (\mathbf{P}\cdot\mathbf{p})}{m_1 m_2 (m_1+m_2)^3}
	+ \frac{(m_1^2-m_1 m_2+m_2^2) (\mathbf{p})^2}{2 m_1^2 m_2^2 (m_1+m_2)^2}
	\Big) (\mathbf{P}\cdot\mathbf{x})
  \nonumber \\ & \quad \quad
      + \frac{G (3 m_1+m_2) (m_1+3 m_2) (\mathbf{P}\cdot\mathbf{x})}{4 (m_1+m_2)^3 \sqrt{(\mathbf{x}\cdot\mathbf{x})}}
      \Bigg)
  \nonumber \\ & \quad
    + P^k\Bigg(
      \Big( \frac{(\mathbf{P}\cdot\mathbf{p})}{2 (m_1+m_2)^4}
	+ \frac{(-m_1+m_2) (\mathbf{p})^2}{4 m_1 m_2 (m_1+m_2)^3}
	\Big) (\mathbf{P}\cdot\mathbf{x})
  \nonumber \\ & \quad \quad
      + \frac{G m_1 (m_1-m_2) m_2 (\mathbf{P}\cdot\mathbf{x})}{4 (m_1+m_2)^4\sqrt{(\mathbf{x}\cdot\mathbf{x})}}
      - \frac{5 G (\mathbf{p}\cdot\mathbf{x})}{4 (m_1+m_2)\sqrt{(\mathbf{x}\cdot\mathbf{x})}}
      \Bigg)
  \nonumber \\ & \quad
    + x^k\Bigg(
      \frac{(-m_1^2+m_2^2) (\mathbf{p})^4}{8 m_1^3 m_2^3}
      + (\mathbf{P}\cdot\mathbf{p}) \Big(
	  \frac{(\mathbf{P})^2}{8 (m_1+m_2)^4}
	+ \frac{(\mathbf{p})^2}{2 m_1 m_2 (m_1+m_2)^2}
      \Big)
  \nonumber \\ & \quad \quad
      + \frac{G m_1 (m_1-m_2) m_2 (\mathbf{P}\cdot\mathbf{x})^2}{2 (m_1+m_2)^4(\mathbf{x}\cdot\mathbf{x})^{3/2}}
      - \frac{G (m_1-m_2)^2 (\mathbf{P}\cdot\mathbf{x}) (\mathbf{p}\cdot\mathbf{x})}{4 (m_1+m_2)^3(\mathbf{x}\cdot\mathbf{x})^{3/2}}
      + \frac{G (-m_1+m_2) (\mathbf{p}\cdot\mathbf{x})^2}{4 (m_1+m_2)^2(\mathbf{x}\cdot\mathbf{x})^{3/2}}
  \nonumber \\ & \quad \quad
      + \frac{G^2 m_1 (m_1-m_2) m_2 (m_1^2+m_2^2)}{4 (m_1+m_2)^3 (\mathbf{x}\cdot\mathbf{x})}
      + \frac{G (5 m_1^2+6 m_1 m_2+5 m_2^2) (\mathbf{P}\cdot\mathbf{p})}{4 (m_1+m_2)^3\sqrt{(\mathbf{x}\cdot\mathbf{x})}}
  \nonumber \\ & \quad \quad
      - \frac{G (m_1-m_2) (5 m_1^2+8 m_1 m_2+5 m_2^2) (\mathbf{p})^2}{4 m_1 m_2 (m_1+m_2)^2\sqrt{(\mathbf{x}\cdot\mathbf{x})}}
    \Bigg)
  \Bigg]
\end{align}

%% file: generat2pn.tex
\begin{equation}
g_{\op{point}}
  = \frac{1}{2}\left[
     g_{\op{1PN}}
   + g_{\op{2PN}}
   + \left(1\leftrightarrow 2\right)
   \right]
\end{equation}

\begin{subequations}
\begin{eqnarray}
g_{\op{1PN}}  &= &
  c^{-2}\bigg\{ \bigg[
    \frac{m_2^{2}}{m_1 m^{3}} \parenmacro{\mathbf{n}_{12}\cdot\mathbf{p}_{1}} p_{1}^{2} r_{12}
  + \frac{m_2^{2}}{m_1 m^{3}} \parenmacro{\mathbf{n}_{12}\cdot\mathbf{p}_{2}} p_{1}^{2} r_{12} 
  \\
  &&
  - \frac{2m_2 -m}{m^{3}} \parenmacro{\mathbf{n}_{12}\cdot\mathbf{p}_{1}} \parenmacro{\mathbf{p}_{1}\cdot\mathbf{p}_{2}} r_{12} \bigg]
  - G \frac{\mu(2m_2- m)}{m} \parenmacro{\mathbf{n}_{12}\cdot\mathbf{p}_{1}} \bigg\} \nonumber
\end{eqnarray}
\begin{eqnarray}
g_{\op{2PN}} & = &
  c^{-4}\bigg\{\bigg[
  - \frac{m_2^{2}(m_2+2 m_1)}{4 m_1^{3} m^{4}} \parenmacro{\mathbf{n}_{12}\cdot\mathbf{p}_{1}} \left(p_{1}^{2} \right)^{2}\ r_{12}
  - \frac{m_2^{2}(m_2+2 m_1)}{4 m_1^{3} m^{4}} \parenmacro{\mathbf{n}_{12}\cdot\mathbf{p}_{2}} \left(p_{1}^{2} \right)^{2}\ r_{12} 
  \nonumber\\
  &&
  + \frac{2m_2+\mu- m}{2 m_1^{2}\ m^{3}} \parenmacro{\mathbf{n}_{12}\cdot\mathbf{p}_{1}} \parenmacro{\mathbf{p}_{1}\cdot\mathbf{p}_{2}}
    p_{1}^{2} r_{12}
  + \frac{2m_2+\mu- m}{2 m_1^{2}\ m^{3}} \parenmacro{\mathbf{n}_{12}\cdot\mathbf{p}_{2}} \parenmacro{\mathbf{p}_{1}\cdot\mathbf{p}_{2}} 
    p_{1}^{2} r_{12} 
  \nonumber\\
  &&
  - \frac{2m_2- m}{2\mu\ m^{4}} \parenmacro{\mathbf{n}_{12}\cdot\mathbf{p}_{1}} \parenmacro{\mathbf{p}_{1}\cdot\mathbf{p}_{2}}^{2} r_{12}
  - \frac{2m_2- m}{4\mu\ m^{4}} \parenmacro{\mathbf{n}_{12}\cdot\mathbf{p}_{1}} p_{2}^{2} p_{1}^{2} r_{12} \bigg]
  \nonumber\\
  && 
  - G \bigg[
      \frac{m_2^{2}(m_2- m_1)}{4 m^{4}} \parenmacro{\mathbf{n}_{12}\cdot\mathbf{p}_{1}}^{3}
    + \frac{m_1(-5m_2+6\ \mu+ m)}{4 m^{3}} \parenmacro{\mathbf{n}_{12}\cdot\mathbf{p}_{2}} \parenmacro{\mathbf{n}_{12}\cdot\mathbf{p}_{1}}^{2}
    \nonumber\\
    && 
    + \frac{2\ \mu(3\mu-8 m)- 5m_2(\mu-2 m)}{4 m_1 m^{2}} \parenmacro{\mathbf{n}_{12}\cdot\mathbf{p}_{1}}\ p_{1}^{2}
    \nonumber\\
    && 
    + \frac{6\mu^{2}-5m_2\mu-16 m\mu+5 m^{2}+5\ m_2 m}{4 m_1 m^{2}} \parenmacro{\mathbf{n}_{12}\cdot\mathbf{p}_{2}}\ p_{1}^{2} 
    \nonumber\\
    && 
    + \frac{6 m^{2}-17\ m_2 m-3\mu m+6m_2\mu}{2m^{3}} \parenmacro{\mathbf{n}_{12}\cdot\mathbf{p}_{1}} \parenmacro{\mathbf{p}_{1}\cdot\mathbf{p}_{2}} 
  \bigg] \nonumber\\
  && 
  + G^{2} \frac{\mu(2\mu- m)( m-2m_2)}{2\ m}\frac{\parenmacro{\mathbf{n}_{12}\cdot\mathbf{p}_{1}}}{r_{12}} 
  \bigg\},
\end{eqnarray}
\end{subequations}